\documentclass[journal,transmag]{IEEEtran}
\usepackage{cite}
\usepackage{amsmath,amssymb,amsfonts}
\usepackage{algorithmic}
\usepackage{graphicx}
\usepackage{textcomp}
\usepackage{libertine}
\usepackage{array}
\usepackage{mdwmath}
\usepackage{mdwtab}
\usepackage{eqparbox}
\usepackage{url}
\usepackage{amsmath}
\usepackage{balance}
\usepackage{blindtext}
\usepackage[T1]{fontenc}
\usepackage[utf8]{inputenc}
\usepackage{CJKutf8}
\usepackage{amsfonts}
\usepackage{setspace,lipsum}
\usepackage{caption}
\usepackage{subcaption}
\usepackage{xcolor}

\ifCLASSINFOpdf
  
\else
  
\fi

\def\BibTeX{{\rm B\kern-.05em{\sc i\kern-.025em b}\kern-.08em
  T\kern-.1667em\lower.7ex\hbox{E}\kern-.125emX}}

\begin{document}
%
\title{Speaker Identification from emotional and noisy speech data using learned voice segregation and Speech VGG}


\author{\IEEEauthorblockN{
Shibani Hamsa\IEEEauthorrefmark{1}, Ismail Shahin\IEEEauthorrefmark{2}, ~\IEEEmembership{{Member,~IEEE}}
 Youssef Iraqi\IEEEauthorrefmark{3}, ~\IEEEmembership{{Senior Member,~IEEE}}\\ Ali Bou Nassif\IEEEauthorrefmark{4}  ~\IEEEmembership{{Member,~IEEE}}, Ernesto Damiani\IEEEauthorrefmark{1},~\IEEEmembership{{Senior Member,~IEEE}}, and Naoufel Werghi\IEEEauthorrefmark{1},~\IEEEmembership{Senior Member,~IEEE}}

\IEEEauthorblockA{\IEEEauthorrefmark{1}Center for Cyber-Physical Systems (C2PS), Dept. of ECE, Khalifa University, Abu Dhabi, UAE.}
\IEEEauthorblockA{\IEEEauthorrefmark{2} Dept. of Electrical Engineering, University of Sharjah, Sharjah, UAE.
\IEEEauthorblockA{\IEEEauthorrefmark{3}School of Computer Science, Mohammed VI Polytechnic University, Morocco}
\IEEEauthorblockA{\IEEEauthorrefmark{4} Dept. of Computer Engineering, University of Sharjah, UAE.}}

\thanks{
Corresponding author: Shibani Hamsa (email: 100050116@ku.ac.ae).}

\markboth{IEEE Transactions on human-machine systems}%
{Shell \MakeLowercase{\textit{et al.}}: Bare Demo of IEEEtran.cls for IEEE Transactions on Magnetics Journals}}
%



\IEEEtitleabstractindextext{%
\begin{abstract}
Speech signals are subjected to more acoustic interference and emotional factors than other signals.  Noisy emotion-riddled speech data is a challenge for real-time speech processing applications. It is essential to find an effective way to segregate the dominant signal from other external influences. An ideal system should have the capacity to accurately recognize required auditory events from a complex scene taken in an unfavorable situation. This paper proposes a novel approach to speaker identification in unfavorable conditions such as emotion and interference using a pre-trained Deep Neural Network mask and speech VGG. The proposed model obtained superior performance over the recent literature in English and Arabic emotional speech data and reported an average speaker identification rate of  85.2\%, 87.0\%, and 86.6\% using the Ryerson audio-visual dataset (RAVDESS), speech under simulated and actual stress (SUSAS) dataset and Emirati-accented Speech dataset (ESD) respectively.

\end{abstract}

}

\maketitle

\IEEEdisplaynontitleabstractindextext

%
\IEEEpeerreviewmaketitle

\section{Introduction}
\label{Section:1}
The human auditory system can handle complex auditory scenes and is efficient enough in precisely distinguishing the various auditory events. There are many different terms to describe this phenomenon, such as segregation by type or frequency, but they all share one important quality: they are efficient. The human ear can only hear so many different types of sound at once but can still distinguish them with great accuracy. A well-constructed soundscape that could be described as perfectly balanced and harmonious. The challenges most of the real-time human-machine interacting audio systems face when handling complex auditory scenes are discussed in this work. The auditory scene is a series of sounds of which the signal-to-noise ratio varies. We must process this, as our auditory experience does not provide these signals in isolation, and isolated sounds are non-existent in the real world—the cocktail party effect \cite{arons1992review} demonstrates this problem, where we attempt to segregate the necessary speech signals from all the other noises in the auditory field, at a party full of noise and chatter \cite{zhang2021predicting}. The goal of this work is to propose a way to tackle these challenges by using machine learning and data mining techniques, where some pre-processing steps are performed to generate appropriate feature representation for a collaborative filtering/learning algorithm, which subsequently leads to the generation of a new sound event classifier. Deep Learning is a form of machine learning using artificial neural networks to model high-level abstractions in data \cite{xin2018machine} \cite{hamsa2020emotion}. The closed-form solution \cite{ho2007accurate} allows one to learn the structure of data without requiring costly supervised or unsupervised pre-processing. Deep Neural Networks differ from standard approaches in how they are learned, which means that rather than relying on hand crafted features \cite{hamsa2021enhanced}, the deep neural network can simply be exposed to lots and lots of data and it will learn features automatically. In this work, we have used machine learning models for dominant signal extraction and speaker identification from  emotional and noisy speech data.

In recent years, we have seen many real-time human-machine interactions that mainly focus on audio. In this work, we have focused on designing and implementing a model suitable for identifying the unknown speaker in emotional and noisy real application situations. The proposed speaker identification model, designed in the deep learning platform, has been evaluated in noisy, stressful, and emotionally challenging talking environments to ensure the system's robustness in real applications \cite{tamada2020noise}. The proposed model achieves the same or better than the best previous state-of-the-art models in most of the evaluation metrics. In addition, it is robust to various acoustic distortions and interference. Finally, we evaluated the effectiveness of the proposed system through its performance on various evaluation metrics.

The rest of the  paper is organized as follows. A literature review is given in Section \ref{Section: 2}. System description is explained in Section \ref{Section: 3}.  Experiment results are described in Section \ref{Section: 4}, and finally conclusion is given in Section \ref{Section: 5}.

\section{LITERATURE REVIEW}
\label{Section: 2}

The Auditory Scene Analysis (ASA) is based on a theory that describes how the brain processes sounds as a result of neural networks. The term "ASA" has been used in the literature to refer to several fields, including music, speech, prosody, and language \cite{bregman1998psychological}. 
Acoustic scene analysis is a method used to extract acoustic environmental information. The technique attempts to model soundscape as a series of layers, each representing the temporal variations of specific properties (e.g., intensity, fundamental frequency, etc.). Acoustic scenes are then analyzed using spectral clustering which finds "typical" or "normal" patterns within an acoustic scene that can be used for recognition purposes \cite{bregman1998psychological}. Nonetheless, the concept of ASA has received considerable attention in the years following its introduction. The ASA principle was applied to a much broader spectrum of auditory stimuli including non-speech sounds. In addition, new theories based on Computational Auditory Scene Analysis (CASA) were developed to explain how humans extract speech and music signals from noise or reverberation \cite{brown1994computational}. The field of computational study known as "speech segregation" refers to the processing of signals in which one or more sources are generating sounds that are being detected by a microphone. One goal is to separate the speech signal from noise signals since any noise can interfere with making out what someone's saying. \cite{wang2018supervised}. For a single microphone, the speech source is located at the microphone's center. Since we cannot place a second microphone close to the location of the original one, we cannot differentiate between the two sources \cite{cardoso1997infomax}. Blind source separation minimizes errors caused by noise and other unwanted noises when compared to conventional sound separating methods. The main idea behind this method is to use two microphones and cancel out signals from both  \cite{cardoso1997infomax}. In speech processing, in recent years, source separation methods have been extensively investigated to obtain clean speech signals from a mixture of multiple talkers. This is because the dynamic range of a real-world signal (in particular human speech) is very wide. The use of an overlapping independent set of sources has been proposed as an effective method for capturing the various sources and their respective signals. \cite{antoni2013study}.

The CASA systems come in two types: data-driven and prediction-driven  \cite{brown2005separation}. The data-driven method is based only on the input signal attribute. The system that uses this architecture is also based on the input signal features. It is called bottom-up since they are built at a higher level even though the data collected from the signals are at a low level. On the contrary, the prediction-driven approach defines the top-down system \cite{shao2010computational}. This architecture is based on the predictions of future outputs. In other words, this system predicts the next attribute of a signal. Therefore, top-down approaches are based on high-level features compared to bottom-up approaches, which are based on low-level features. The data-driven approach has more stability but less adaptability and flexibility than the prediction-driven approach, which has less stability but more adaptability and flexibility in many cases. 

 Meddis \cite{meddis1997unitary} and O'Mard prepared the most efficient pitch estimation models. Since they used multi-channel models, they are not suitable for speech separation applications such as hearing aids which require a single channel pitch estimation algorithm. The ITU-T G.1204 and G.1205 standards require that hearing aids \cite{haakansson1985bone} be capable of separating speakers in a room. This can be done by identifying the dominant speaker among several speakers. The pitch estimation algorithm provides a score for each speaker's contribution to the signal mixed in the acoustic domain coupled with speech recognition to contribute to the identification of the dominant speaker.

In CASA systems, time-frequency decomposition is performed using auditory filters, whose bandwidth increases quasi-logarithmic concerning center frequencies. Since the effective speech separation algorithm or one of its derivatives is defined as a 2D spectral ratio of two time-frequency signals, it cannot be given by some simple formula. The main difficulty in these algorithms is to estimate the parameters for which the resulting signal is most suited for subsequent Minimum Mean Square Error (MMSE) filtering and vice versa. These filters are derived from the psychophysical observations of the auditory periphery. An auditory filter bank is used to imitate cochlear filtering. There are two such filter banks: the Gamma-tone filter bank and the Short-time Fourier transform (STFT) based filter bank \cite{hamsa2021enhanced}. The STFT filter bank is more efficient as it utilizes the high-resolution capabilities of Digital Signal Processing (DSP) hardware \cite{mahmoodzadeh2010single}. Hamsa \textit{et al.} proposed and implemented a wavelet packet transform (WPT) based filter bank for segregating noise and emotional speech data \cite{hamsa2020emotion}.

 Emotion attribute projection (EAP) and linear fusion were used by Bao \textit{et al.} \cite{bao2007emotion} to analyze speech, design a recognition system for speaker identification in emotional speaking conditions, and validate the system through evaluation of real data. The findings were that linear fusion provided an improvement to EAP based emotion recognizer for mental well-being in emotional speaking condition. Shahin \textit{et al.} focused on improving the performance of techniques for voice identification in emotional speaking conditions \cite{shahin2019emotion}. His studies include improving speaker identification performance based on hand-crafted features such as Hidden Markov Models (HMMs), Second Order Circular Hidden Markov Models (CHMM2s), and Supra-segmental Hidden Markov Models (SPHMMs). Each of these models achieved average speaker identification performance, with the highest being SPHMMs having 69.1\% followed by CHMM2s and HMMs with 66.4\% and 61.4\%, respectively \cite{shahin2020novel}. For improved results, the authors used and assessed a hybrid Gaussian Mixture Model-Deep Neural Network (GMM-DNN) classifier and obtained an average speaker identification rate of 76.8\% \cite{shahin2020novel}. Nassif \textit{et al.} improved the results of the GMM-DNN model by adding a suitable noise reduction pre-processing module based on CASA \cite{nassif2021casa}.

In this paper, we have designed and applied a more coherent and less complex model than the existing models for speech segregation and identification of the unknown speaker in emotional and noisy talking conditions. The proposed algorithm utilizes pre-trained deep learning approaches for speech segregation, feature extraction, and classification.  The state-of-the-art model used onset-offset-based segmentation and classification for dominant voice segregation, in which the pitch of target and interference were assessed. The pitch of the signals is essential as it is used to develop a spectral mask to separate the speech from the noisy part \cite{xiao2017time}. The highlighted contributions of this work are:
\begin{itemize}
\item Blending of pre-trained deep-learning model for CASA and speaker identification in challenging talking conditions have not been presented before to the best of our knowledge.
    \item Learned approach has been used for retrieving original voice information from the emotional and environmentally contaminated input signal.
      \item WPT filter-bank is introduced to reduce the computational complexity associated with the conventional filter-bank models in the recent literature.
      \item Pre-trained speech VGG module is proposed for speaker identification in emotional and noisy challenging talking conditions.
\end{itemize}

 \begin{figure*}[htb]
      \centering
      \includegraphics [scale=.22]{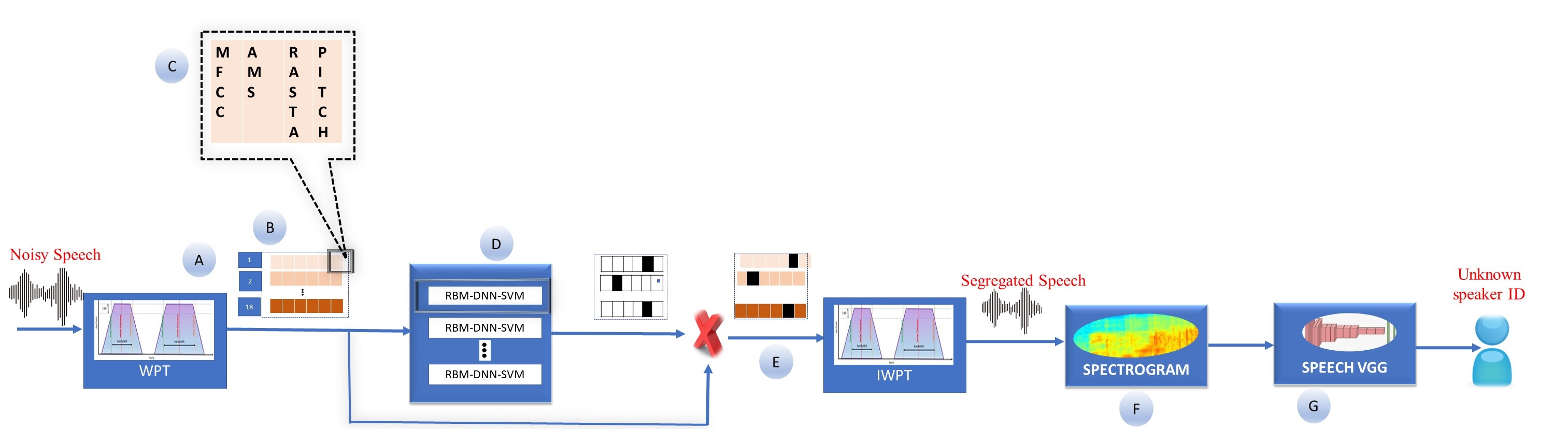}
      \caption{Block schematic of the proposed model. Each block labeled in alphabetical order is been described as subsections in section \ref{Section: 3} at the same order.}
      \label{1_Block}
   \end{figure*}
   
    

\section{SYSTEM DESCRIPTION}
As depicted in Fig \ref{1_Block}, the block diagram of the proposed system encompasses a 7-staged framework for speaker identification in challenging talking conditions. First, the input signal is passed through the WPT Cochlear filter-bank. Raw features are extracted from each channel segment are allowed to pass through a hybrid classifier model to estimate the binary T-F mask. The dominant speech signal segregated by means of IBM is re-synthesized and used for final classification. More details about each stage is presented in the following sub-sections.
\label{Section: 3}
\subsection{WPT Cochlear filterbank}
The proposed framework utilizes a novel differentiation interface based on wavelet transform to estimate the spectral transfer function. The smoothing effect of the filter-bank is capable of reducing noise artifacts. Gammatone filter-bank or the Short-Time Fourier Transform (STFT) filter-bank are often used in the greater part of speech segregation and noise cancellation studies. Conversely, based on the latest understanding, the computational cost associated with the Gamma-tone filter-bank is quite high \cite{el2011survey}. STFT-based static extraction always exhibits a conflict between temporal and spectral resolution. In this work, we have used a novel db4-WPT cochlear filter bank model for acoustic energy extraction. This filterbank is applied by computing an approximate Recursive Wavelet Transform on each frequency channel separately using overlapping windows in which neighboring windows overlap by one-third of their window widths. The proposed db4 WPT splits the complex acoustic scene into 18 channels \cite{hamsa2020emotion}.

\subsection{Channel Segmentation}
In this research, complex auditory scenes at a sampling rate of 16 kHz are separated into individual components by means of a WPT filter-bank. it can still be segmented and analyzed as individual channels of audio data resulting in a T-F representation termed cochleagram. These segments are only 20ms long so they can be appropriately analyzed in frequency space as well as pitch space.
\subsection{Raw features Extraction}
   The extraction of segments that are good for detecting linguistic material and removing the rest is the concept behind noise suppression or speech enhancement.  However, In our work, we need to consider the emotional component, along with the linguistic quality of the sound from the noise and interference signals to achieve improved efficiency in emotional, conversational environments. Raw features such as MFCC (Mel frequency cepstral coefficient) \cite{hamsa2020emotion}, Amplitude modulation spectrogram (AMS), RASTA-PLP, and pitch are extracted from each segment to isolate the speech and emotional information from the disruptive portion of the speech and to improve the quality of classification. These three types of features will be described next. 
\subsubsection{Mel Frequency Cepstral coefficients}
Speakers' voices contain many different frequencies, that is why voice recognition systems require too much data. The Mel Frequency Cepstral Coefficient (MFCC) is the fundamental frequency composition of speech in the unit Hz. It is also used to identify speakers from their voice recordings, primarily in forensic speech processing. 

The formula for translation from frequency to Mel scale is \cite{davis1980comparison}:
\begin{equation}
    M(f)= 1125\ln(1+\frac{f}{100})
\end{equation}
Where $f$ and $M(f)$ represent the frequency and Mel-frequency, respectively. The implementation steps are as follows \cite{davis1980comparison}:

\begin{itemize}
\item Frame the signal into short frames.
    \item For every frame, compute the periodogram estimate of the power spectrum.
    \item Implement the Mel filterbank to the power spectra, then add the energy in each filter.
    \item Take the logarithm of all filterbank energies.  
    \item Take the DCT (Discrete Cosine Transform) of the log filterbank energies.
   
\end{itemize}
\subsubsection{Amplitude Modulation Spectrum}
The input emotional and environmentally contaminated speech data can be split into short segments of 20ms duration. Spectral information is extracted from each short segment, and the resultant magnitude and phase information are recorded as a matrix:
\begin{equation}
   DWPT\underset{m}(s(n))=S_i(m,k) 
\end{equation}
where $s(n)$ is the wide-band speech signal and $S(m,k)$ is its time \textit{vs} frequency representation. $m$ and $k$ denotes the time and frequency respectively.

The T-F domain signal $S_i(m,k)$ encapsulates the message part and the carrier part of the speech signal and are represented as $Message(m,k)$ and $Carrier(m,k)$ respectively [28]:
\begin{equation}
     S_i(m,k)= Message_i(m,k) Carrier_i(m,k) 
\end{equation}
 
 In this work, we have used the "Hilbert envelop" detection approach to extract the message part $Message_i(m,k)$, and it is approximated as the magnitude of the complex sub-band \cite{qin2006new}.
\begin{equation}
     Message_i(m,k) \approx |S_i(m,k)|
\end{equation}
The resultant signal is set as,
\begin{equation}
     S_i (k,\omega)  = DFT(  Message_i(m,k))
\end{equation}
 where $\omega $ represents the modulation frequency index. The modulation components outside the range of 1-16 Hz are discarded, and the features vector size is reduced to 38 using Principle component analysis (PCA) to preserve comparability with MFCCs.
\subsubsection{RASTA PLP} 
Acoustic signals can vary significantly in structure, complexity, and type. For a machine learning system to be able to make sense of these signals, it needs a way to extract the most critical variables from each signal and then compare these extracted variables between different acoustic signals. it is necessary for the machine learning system to understand which features are most important in analyzing speech variation over time. Relative SpectrAl Phonetic Likelihood Probabilities (RASTA PLP) 
does just this by finding and extracting the phonetic distance from one acoustic signal and comparing it with other phonetic distances
from other acoustic signals \cite{hermansky1991rasta}. 
It is based on the principle that an edge-detected image of a sound field is similar to the waveforms it reflects, i.e., its phase structure should highlight edges and their relative positions. RASTA PLP feature extraction  can be implemented using the asymmetric discrete cosine transform, which is computationally fast and has been shown to perform well on voiced and unvoiced speech signals \cite{hermansky1991rasta},

 \subsection{Segment classification}

A key stage of speaker identification is the  separation of the emotional component and the linguistic quality of the sound from the noise and interference signals to achieve improved efficiency in emotional, conversational environments. 
Before speaker identification, a  popular way to perform feature enhancement was using computational auditory scene analysis (CASA) based algorithms to perform speech separation. The main goal of CASA is to estimate the ideal binary mask (IBM). A binary  mask is simply a 2D binary array that separates the voice component from the complex auditory scene. An IBM is defined as follows \cite{wang2005ideal}: 
   \begin{equation}
   IBM (t,f)= 
     \begin{cases}
     1, & \text{if}\,
   \text{LSNR (t,f)}>= \text{LC} \\
     0, & \text{otherwise}
     \end{cases}
   \end{equation}
 where LSNR (t, f) corresponds to the specific signal-to-noise ratio of the T-F unit in time $t$ and frequency $f$ and LC is the local threshold. IBM reduces noise cancellation as a simple binary classification problem, which can be solved by supervised learning algorithms.


In this research, we adopted a machine learning approach for estimating the binary mask. Our method is inspired by a classification-based speech segregation model introduced by wang et al.\cite{wang2015speech}. They employed a typical 64-channel gammatone filter bank to simulate the frequency resolution of the human auditory system. A Deep Belief Network (DBN) model was proposed to learn more linearly separable features from raw attributes. Linear SVMs trained on a mixture of characteristics classify each segment in the cochleagram as signal dominant or not. To accommodate the variations in spectral characteristics across frequencies, each subband classifier is trained separately based on the IBM ground truth.

We adopted a similar framework for binary classification as introduced in \cite{wang2015speech}. However, we have replaced the 64 channel gammatone filter bank in their model with an 18 channel WPT filterbank. In doing so, we bring the number of 
subband classifiers from 64 down to 18, hence reducing significantly the  computational complexity. 
In our work,  the input speech signal is therefore routed through an 18-channel WPT filter bank with center frequencies of about 50 Hz to 4000 Hz on an equivalent rectangular bandwidth rate scale, and the Cochleagrams are created using these 18 channels. 

Pre-trained DNN using Restricted Boltzmann Machines (RBM) and SVMs are being used for mask estimation and prediction based on speaking environments. Each channel has a hybrid RBM-DNN-SVM classifier model. An unsupervised pre-trained RBM is used to initialize the multilayer perceptrons (MLPs) to address the challenges associated with training.
 The weights from the last hidden layer to the output layer of MLPs would primarily characterize a linear classifier. Hence, the last hidden layer responses are more amenable to linear classification."Wang \textit{et al.} \cite{wang2015speech} reported that concatenating the learned features with the raw features could result in HIT-FA improvements in practice. We followed this directive by selecting the deepest  hidden layer activations as the learned features. Then, SVMs utilizing both raw and learned features were used for binary classification." Each channel is independently trained  using 100 male and female voice samples from the TIMIT dataset combined with 100 environmental noises at 0 dB. The classification results from the 18 subband classifiers yield an estimated IBM.  
 
 Based on the above, we recapitulate our IBM estimation process in the following steps: 
\begin{enumerate}
   \item Input signal is passed through an 18 channel WPT filter bank.
   \item Cochleagrams are formulated in each channel.
    \item Raw features AMS, RASTA-PLP, and MFCC are extracted.
    \item Raw features are used to train the first RBM network in the stack.
    \item The DNN is initialized with weights obtained from the RBM stack trained in layer-wise fashion.
    \item The whole network is then fine-tuned using a back propagation algorithm with a logistic output layer. 
   \item The last hidden layer activations are chosen as the learned features. 
    \item Linear classifier (SVM) is trained on a combination of both refined and learned attributes to predict IBM.
\end{enumerate}

\subsection{Re-synthesis of the segregated speech signal}
Reconstruction of the segregated speech signal from the noisy input signal after estimation is illustrated in Fig \ref{1_Block}(E). The magnitude of the noisy input is multiplied by the estimated mask in the spectral domain. The phase of the mixture is appended with the magnitude during the spectral to time-domain conversion.  


\subsection{Spectrogram Extraction}
In this part, the system  extracts the spectrogram from each of the segregated speech signals. The spectrogram is an image that represents the spectrum of frequencies of an audio signal over time. The brighter color in the spectrogram indicates that  the sound is heavily concentrated around those specific frequencies, while the dark region point to empty/dead sound. In this way, we can revamp  complex speech signals as a 2D image. This is where the power of spectrograms comes into play for various Machine learning/Deep Learning models.

\subsection{Classification using SpeechVGG}
 The classifier model used in this work is a transfer learning approach, inspired from the conventional VGG-16 ConvNET  \cite{qassim2018compressed}. Beckmann and Keggler \cite{beckmann2019speech} proposed a speech VGG framework suitable for various speech processing applications using the TIMIT dataset in the recent literature. However, the speech VGG modeled here in this work is an end-to-end speaker identification classifier network trained on spectrograms extracted from emotional speech corpus. The proposed topology of the system consists of 5 blocks of stacked convolution layers followed by a ReLU activation layer, and a max-pooling layer.  The result of the convolution layers is refined through two linear fully connected layers and a softmax output layer. A max-pooling layer subsequently follows this stack. To the best of our knowledge, this is the first work that utilizes a VGG-16 based transfer learning approach for speaker identification and experimental evaluation in emotional and noisy challenging talking conditions.

\section{Experiments} \label{Section: 4}
To assess and validate our framework, we conducted a series of experiments on two public and private datasets. These experimentation aims to 
\begin{itemize}
  
  \item Evaluate the performance of our system with competitive state of the art methods 

  \item Study the effect of integrating CASA on our system.

  \item  Statistical analysis of the system performance to check the significance of the results compared to the recent literature.

  \item To compare the model's performance in noisy and normal talking conditions.
\end{itemize}

In the following, we will give a brief description of these datasets. Afterward, we will elaborate on the different experiments. 

\subsection{Speech Datasets}
We used a group of datasets encompassing  English and Emirati accented Arabic languages composed of  the "Speech Under Simulated and Actual Stress (SUSAS) dataset \cite{hansen1997getting}, the  Ryerson Audio-Visual Database of Speech and Song (RAVDEES)" \cite{livingstone2018ryerson} datasets, and the Emirati speakers dataset.

\subsubsection{SUSAS Dataset}
This dataset comprises of five domains with an array of stress and emotional features. The two main domains used are simulated and actual domains. The former involves simulated speech under stress, and the latter is actual speech under stress. 19 male and 13 female speakers between 22 and 76 years uttered more than 16,000 words in this work. The speech samples were sampled by 8 kHz using an Analog to Digital (A/D) converter. The speech tokens were pre-emphasized and separated into frames of 20ms, each overlapping 31.25\% between frames. These speech signals were implemented to a 30ms Hamming window  \cite{hansen1997getting}.

\subsubsection{RAVDESS Dataset}
The RAVDESS \cite{livingstone2018ryerson} includes a verified multimodal database of song and speech. This database consists of 12 professional actors and 12 professional actresses, each pronouncing lexically matched phrases in a neutral North American accent. The speech and songs contain happy, disgust, sad, fearful, angry, and neutral emotions. We have taken 60 trials consisting of 1,440 files by each speaker.

\subsubsection{ESD}
There are 50 Emirati speakers, 25 female and 25 male speakers between 14 and 55 years. They were asked to utter eight public Emirati utterances that are comprehensively utilized in the United Arab Emirates in Emirati emphasized speech database (Arabic). They spoke the eight sentences with each of the mentioned emotions – sad, neutral, fear, disgust, angry, and happy in 2 – 5 seconds \cite{shahin2020novel}.

\subsection{Model Training}
The spectrograms in the log scale are quantified for each segment of sample size 256 with 128 overlaps. The obtained speech segments are interpolated to 1024ms extended frames using zero padding to mitigate the effect of changes in talking time. SpeechVGG was trained using categorical cross-entropy for $25$ epochs through an ADAM optimizer \cite{zhang2018improved}. The learning rate used is $4\times10^{-5}$.



\begin{figure*}[htbp]
      \centering
      \includegraphics [scale=0.90]{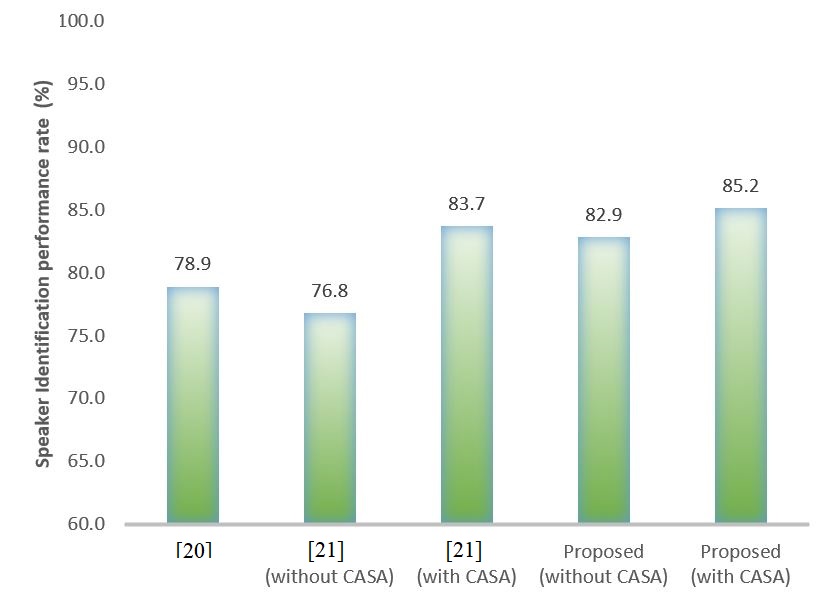}
      \caption{Performance evaluation using RAVDESS.} 
      \label{chart}
   \end{figure*}
   
  \begin{figure*}[htb]
     \centering
     \begin{subfigure}[b]{0.85\textwidth}
         \centering
         \includegraphics[width=\textwidth]{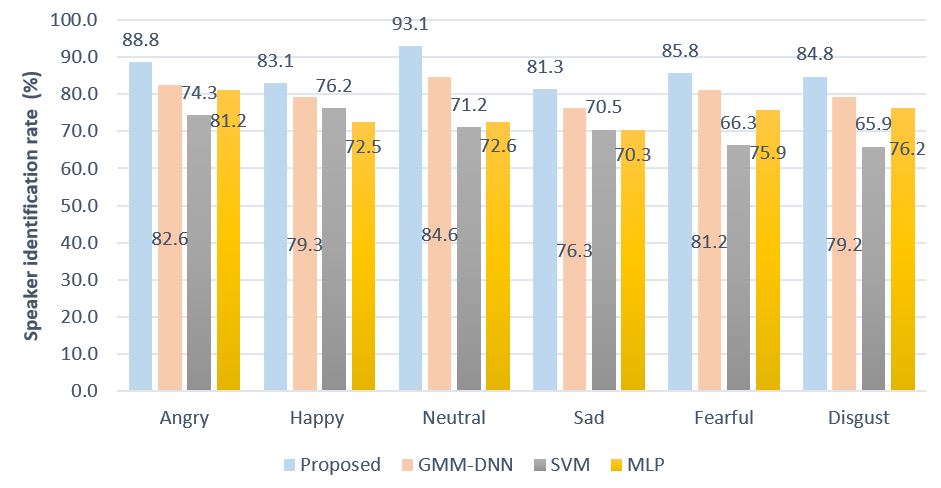}
         \caption{without noises}
         
     \end{subfigure}
     \hfill
     \begin{subfigure}[b]{0.85\textwidth}
         \centering
         \includegraphics[width=\textwidth]{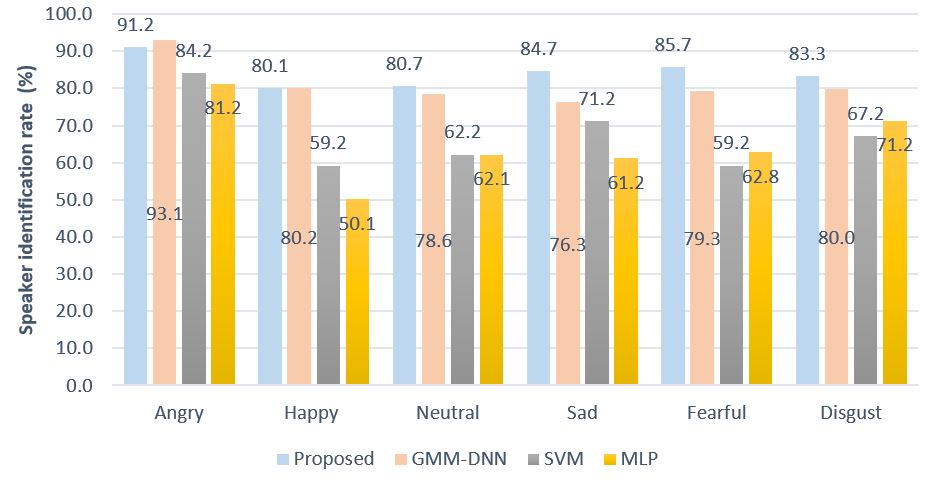}
         \caption{with noises.}
     
     \end{subfigure}
      \caption{Speaker identification performance using RAVDESS dataset}
        \label{experiment1}
\end{figure*}


\subsection{Evaluation Metrics}
To evaluate the proposed technique, speaker identification performance is compared with competitive methods in the recent literature. Speaker identification performance can be defined as:
  \begin{equation}
      \text{SID}=\frac{\text{No.\, of\, times\, speakers\, identified\, correctly}}{\text{Total \,number\, of\, trials}}\times 100 \%
  \end{equation}
  The other evaluation metrics considered for evaluation are listed in this section.
\subsubsection{Statistical Analysis}
Student-$t$ distribution test is carried out to assess the statistical significance of the obtained results. The student-$t$ distribution of two sequences of same length can be described as,
 \begin{equation}
   t_1\,_2=\frac{\Bar{X_1}-\Bar{X_2}}{SD_{pooled}}
   \end{equation}
     \begin{equation}
        SD_{pooled}=\sqrt{\frac{{(SD_1)^2}+{(SD_2)^2}}{2}}
  \end{equation}
  where $\Bar{X_1}$, $\Bar{X_2}$ are the mean and $SD_1$ and $SD_2$ are the standard deviation of the two sequences respectively.
  
  \subsubsection{Precision}
Precision is the ratio of correctly predicted positive observations to the total predicted positive observations \cite{sokolova2006beyond}: \begin{equation}
       Precision = \frac{TP}{TP+FP}
   \end{equation}
 \subsubsection{Recall}  
   The recall is the ratio of correctly predicted positive observations to all observations in actual class \cite{sokolova2006beyond}:
    \begin{equation}
     Recall = \frac{TP}{TP+FN}
   \end{equation}
   \subsubsection{F1 Score}
    F1 Score is the weighted average of Precision and Recall \cite{sokolova2006beyond}:
    \begin{equation}
       F1 Score = \frac{2 \times Recall \times Precision}{Recall +Precision}
   \end{equation}
  where $TP$, $TN$, $FP$, and $FN$ are the true positive, true negative, false positive, and false negative values, respectively, computed from the confusion matrix.

\subsection{Evaluation}
\subsubsection{Experiment 1: Evaluation using RAVDESS dataset.}
  
 Fig. \ref{chart} shows the performance evaluation of the proposed model with that of the state-of-the-art models \cite{shahin2020novel} \cite{nassif2021casa}. The results indicate that the proposed model with CASA for noise reduction offers better performance than all other models in the literature using the RAVDESS dataset.  10-fold cross-validation has been applied to assess speaker identification performance, and hypothesis tests have been used to report statistically significant differences. Speaker identification performance of the proposed algorithm is also computed without noise reduction to check the noise susceptibility of the classifier. In this case, the spectrogram of the input speech signal is directly fed into the speech vgg without any pre-processing.  \\ 
  
\begin{table*}[htbp]
\centering
\caption{Statistical significance test.}
\label{tab:t-value}
\begin{tabular}{|l|l|l|l|l|l|l|}
\hline
\textbf{\begin{tabular}[c]{@{}l@{}}Talking\\   condition\end{tabular}} & \textbf{Angry} & \textbf{Happy} & \textbf{Neutral} & \textbf{Sad} & \textbf{Fearful} & \textbf{Disgust} \\ \hline
$\mathbf{t_{proposed,GMM-DNN}}$ & 1.67 &1.68& \textcolor[rgb]{0,0,1}{1.63} & 1.72 & 1.69 & 1.75 \\ \hline
$\mathbf{t_{proposed,SVM}}$ & 1.69 & 1.72 & 1.66 & 1.66 & 1.67 & 1.71 \\ \hline
$\mathbf{t_{proposed,MLP}}$ &1.72 &  1.80& 1.67   & 1.76 & 1.82 & 1.87 \\ \hline
\end{tabular}
\end{table*}

Table \ref{tab:t-value} shows the calculated $t$ values between the proposed method and each of GMM-DNN, SVM, and MLP using the RAVDESS dataset. Each calculated $t$ value is higher than the “tabulated critical value $t_{0.05}$ = 1.645” except for neutral with GMM-DNN and MLP classifiers (indicated in blue color). The results show that the proposed classifier demonstrates significant improvement in recognizing the speakers in almost all the emotional talking conditions except in the neutral one.

 The proposed model is compared with the recent literature's GMM-DNN \cite{shahin2020novel}, SVM, and MLP \cite{nassif2021casa} classifiers. Fig. \ref{experiment1} displays the speaker identification performance in each of the six emotional talking conditions such as angry, happy, neutral, sad, fearful, and disgust. The results obtained in the evaluation phase using normal emotional speech data are depicted in Fig. \ref{experiment1}(a).
 The RAVDESS corpus has been used to evaluate the performance of the proposed technique with other classifiers in challenging talking conditions. The original speech signals from the RAVDESS dataset are used for training. Then the original speech signals mixed with the noise in a ratio of 2:1 are used for evaluation. Fig. \ref{experiment1}(b) demonstrates speaker identification results using the RAVDESS dataset in noisy talking conditions. An average speaker identification rate of 86.16\% and 83.48\% were obtained, respectively, with the standard and noisy signals. The figure shows that the proposed system performance is superior to that of the other models.

\begin{figure*}[htb]
     \centering
     \begin{subfigure}[b]{0.85\textwidth}
         \centering
         \includegraphics[width=\textwidth]{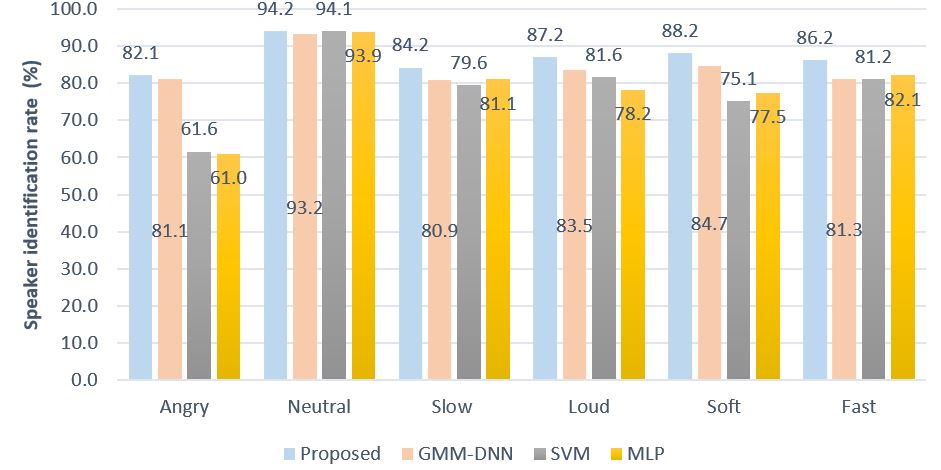}
         \caption{without noises}
         \label{noise susas1}
     \end{subfigure}
     \hfill
     \begin{subfigure}[b]{0.85\textwidth}
         \centering
         \includegraphics[width=\textwidth]{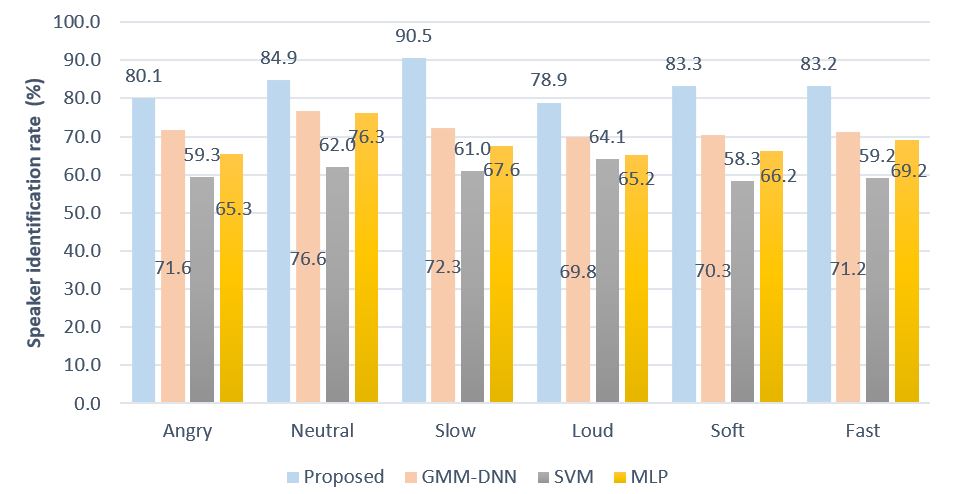}
         \caption{with noises.}
         \label{noise susas2}
     \end{subfigure}
      \caption{Speaker identification performance using the SUSAS dataset}
        \label{noise susas}
\end{figure*}
   
 \begin{table*}[]
\centering
\caption{Evaluation of the proposed classifier model with that of the state-of-the-art models in terms of precision, recall, and F1 score.}
\label{tab:preci}
\begin{tabular}{|l|l|l|l|l|}
\hline
Parameters & Proposed & GMM-DNN \cite{nassif2021casa} & SVM \cite{shahin2020novel}& MLP\cite{shahin2020novel} \\ \hline
Precision  & 0.83      & 0.81     & 0.74 & 0.70 \\ \hline
Recall     & 0.82      & 0.80     & 0.72 & 0.69 \\ \hline
F1 Score   & 0.82      & 0.80     & 0.73 & 0.69 \\ \hline

\end{tabular}
\end{table*}

\subsubsection{Experiment 2: Evaluation using the SUSAS dataset}
Experiment 2 evaluates the proposed system performance in noisy and stressful talking conditions using the SUSAS dataset. The original speech signals from the SUSAS dataset are used for training. The original speech signals mixed with the noise in a ratio of 2:1 are used for testing to evaluate the noise susceptibility of the proposed system in stressful talking conditions. Fig. \ref{noise susas} shows the experiment results. Average speaker identification rates of 87.03\% and 84.27\% were obtained, respectively, with the normal and noisy signals. From the figure, it is apparent that the proposed system performance is better than that of the other techniques in the literature in noisy and stressful talking conditions.

\subsubsection{Experiment 3: Evaluation using ESD dataset}
\begin{figure*}[htpb]
      \centering
\includegraphics   [scale=0.90]{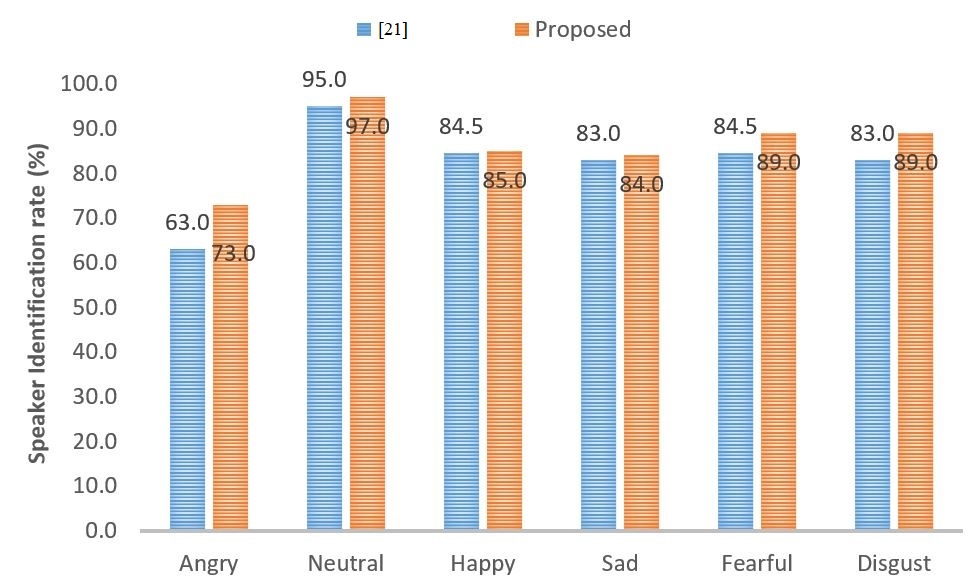}
      \caption {Performance evaluation using ESD.}
      \label{recognition_ESD}
   \end{figure*}
 Fig. \ref{recognition_ESD} compares the speaker identification accuracy of the proposed algorithm with the state-of-the-art model using the ESD dataset. 10-fold cross-validation has been applied to assess speaker identification performance. Higher accuracy is obtained in neutral talking conditions and for other emotional talking conditions, an almost stable response is obtained. The proposed technique offers an average speaker recognition rate of 86.16\% using Emirati emphasized Arabic dataset over the hybrid deep learning classifier model proposed in \cite{nassif2021casa}.

   \subsubsection{Experiment 4}
 
   In this experiment, we have computed the performance interpretations such as precision, recall, and F1 score using the RAVDESS dataset. 
   

   Table \ref{tab:preci} illustrates the performance matrices obtained for the proposed work, GMM-DNN \cite{nassif2021casa}, SVM, and MLP\cite{shahin2020novel} classifier models, respectively. The obtained results depict that the proposed model yields a significant improvement in terms of performance over the other classifiers.

\section{Conclusion}
\label{Section: 5}
The new pipeline has been applied to the emotion in noise speech data in Arabic and English. Deep learning methods are employed for speech segregation and speaker identification in emotional talking conditions. The overall performance of the proposed algorithm is significantly promising when comparing it with other techniques in the recent literature. The inferences collected from the results illustrate the superiority of the proposed model in noisy conditions. To measure the noise immunity of the proposed model, the system has been trained with the pure speech signal and evaluated with noisy speech data. The main limitation of the technique is the computational complexity of the architecture imposed by the multiple deep learning modules. Our future work aims to improve classification efficiency  without increasing computational complexity.

\balance

\bibliographystyle{IEEEtran}
\bibliography{IEEEabrv,bare_jrnl_transmag}

\begin{thebibliography}{10}
\providecommand{\url}[1]{#1}
\csname url@samestyle\endcsname
\providecommand{\newblock}{\relax}
\providecommand{\bibinfo}[2]{#2}
\providecommand{\BIBentrySTDinterwordspacing}{\spaceskip=0pt\relax}
\providecommand{\BIBentryALTinterwordstretchfactor}{4}
\providecommand{\BIBentryALTinterwordspacing}{\spaceskip=\fontdimen2\font plus
\BIBentryALTinterwordstretchfactor\fontdimen3\font minus
  \fontdimen4\font\relax}
\providecommand{\BIBforeignlanguage}[2]{{%
\expandafter\ifx\csname l@#1\endcsname\relax
\typeout{** WARNING: IEEEtran.bst: No hyphenation pattern has been}%
\typeout{** loaded for the language `#1'. Using the pattern for}%
\typeout{** the default language instead.}%
\else
\language=\csname l@#1\endcsname
\fi
#2}}
\providecommand{\BIBdecl}{\relax}
\BIBdecl

\bibitem{arons1992review}
B.~Arons, ``A review of the cocktail party effect,'' \emph{Journal of the
  American Voice I/O Society}, vol.~12, no.~7, pp. 35--50, 1992.

\bibitem{zhang2021predicting}
R.~Zhang, Z.~Wang, Z.~Huang, L.~Li, and M.~Zheng, ``Predicting emotion
  reactions for human--computer conversation: A variational approach,''
  \emph{IEEE Transactions on Human-Machine Systems}, 2021.

\bibitem{xin2018machine}
Y.~Xin, L.~Kong, Z.~Liu, Y.~Chen, Y.~Li, H.~Zhu, M.~Gao, H.~Hou, and C.~Wang,
  ``Machine learning and deep learning methods for cybersecurity,'' \emph{Ieee
  access}, vol.~6, pp. 35\,365--35\,381, 2018.

\bibitem{hamsa2020emotion}
S.~Hamsa, I.~Shahin, Y.~Iraqi, and N.~Werghi, ``Emotion recognition from speech
  using wavelet packet transform cochlear filter bank and random forest
  classifier,'' \emph{IEEE Access}, vol.~8, pp. 96\,994--97\,006, 2020.

\bibitem{ho2007accurate}
K.~Ho and M.~Sun, ``An accurate algebraic closed-form solution for energy-based
  source localization,'' \emph{IEEE transactions on audio, speech, and language
  processing}, vol.~15, no.~8, pp. 2542--2550, 2007.

\bibitem{hamsa2021enhanced}
S.~Hamsa, Y.~Iraqi, I.~Shahin, and N.~Werghi, ``An enhanced emotion recognition
  algorithm using pitch correlogram, deep sparse matrix representation and
  random forest classifier,'' \emph{IEEE Access}, 2021.

\bibitem{tamada2020noise}
D.~Tamada, ``Noise and artifact reduction for {MRI} using deep learning,''
  \emph{arXiv preprint arXiv:2002.12889}, 2020.

\bibitem{bregman1998psychological}
A.~S. Bregman, ``Psychological data and computational {ASA},'' in
  \emph{Computational auditory scene analysis}.\hskip 1em plus 0.5em minus
  0.4em\relax L. Erlbaum Associates Inc., 1998, pp. 1--12.

\bibitem{brown1994computational}
G.~J. Brown and M.~Cooke, ``Computational auditory scene analysis,''
  \emph{Computer speech and language}, vol.~8, no.~4, pp. 297--336, 1994.

\bibitem{wang2018supervised}
D.~Wang and J.~Chen, ``Supervised speech separation based on deep learning: An
  overview,'' \emph{IEEE/ACM Transactions on Audio, Speech, and Language
  Processing}, vol.~26, no.~10, pp. 1702--1726, 2018.

\bibitem{cardoso1997infomax}
J.-F. Cardoso, ``Infomax and maximum likelihood for blind source separation,''
  \emph{IEEE Signal processing letters}, vol.~4, no.~4, pp. 112--114, 1997.

\bibitem{antoni2013study}
J.~Antoni and S.~Chauhan, ``A study and extension of second-order blind source
  separation to operational modal analysis,'' \emph{Journal of Sound and
  Vibration}, vol. 332, no.~4, pp. 1079--1106, 2013.

\bibitem{brown2005separation}
G.~J. Brown and D.~Wang, ``Separation of speech by computational auditory scene
  analysis,'' in \emph{Speech enhancement}.\hskip 1em plus 0.5em minus
  0.4em\relax Springer, 2005, pp. 371--402.

\bibitem{shao2010computational}
Y.~Shao, S.~Srinivasan, Z.~Jin, and D.~Wang, ``A computational auditory scene
  analysis system for speech segregation and robust speech recognition,''
  \emph{Computer Speech \& Language}, vol.~24, no.~1, pp. 77--93, 2010.

\bibitem{meddis1997unitary}
R.~Meddis and L.~O’Mard, ``A unitary model of pitch perception,'' \emph{The
  Journal of the Acoustical Society of America}, vol. 102, no.~3, pp.
  1811--1820, 1997.

\bibitem{haakansson1985bone}
B.~H{\aa}kansson, A.~Tjellstr{\"o}m, U.~Rosenhall, and P.~Carlsson, ``The
  bone-anchored hearing aid: principal design and a psychoacoustical
  evaluation,'' \emph{Acta oto-laryngologica}, vol. 100, no. 3-4, pp. 229--239,
  1985.

\bibitem{mahmoodzadeh2010single}
A.~Mahmoodzadeh, H.~R. Abutalebi, H.~Soltanian-Zadeh, and H.~Sheikhzadeh,
  ``Single channel speech separation with a frame-based pitch range estimation
  method in modulation frequency,'' in \emph{2010 5th International Symposium
  on Telecommunications}.\hskip 1em plus 0.5em minus 0.4em\relax IEEE, 2010,
  pp. 609--613.

\bibitem{bao2007emotion}
H.~Bao, M.-X. Xu, and T.~F. Zheng, ``Emotion attribute projection for speaker
  recognition on emotional speech,'' in \emph{Eighth Annual Conference of the
  International Speech Communication Association}, 2007.

\bibitem{shahin2019emotion}
I.~Shahin, A.~B. Nassif, and S.~Hamsa, ``Emotion recognition using hybrid
  gaussian mixture model and deep neural network,'' \emph{IEEE Access}, vol.~7,
  pp. 26\,777--26\,787, 2019.

\bibitem{shahin2020novel}
------, ``Novel cascaded {Gaussian} mixture model-deep neural network
  classifier for speaker identification in emotional talking environments,''
  \emph{Neural Computing and Applications}, vol.~32, no.~7, pp. 2575--2587,
  2020.

\bibitem{nassif2021casa}
A.~B. Nassif, I.~Shahin, S.~Hamsa, N.~Nemmour, and K.~Hirose, ``{CASA}-based
  speaker identification using cascaded {GMM-CNN} classifier in noisy and
  emotional talking conditions,'' \emph{Applied Soft Computing}, vol. 103, p.
  107141, 2021.

\bibitem{xiao2017time}
X.~Xiao, S.~Zhao, D.~L. Jones, E.~S. Chng, and H.~Li, ``On time-frequency mask
  estimation for {MVDR} beamforming with application in robust speech
  recognition,'' in \emph{2017 IEEE International Conference on Acoustics,
  Speech and Signal Processing (ICASSP)}.\hskip 1em plus 0.5em minus
  0.4em\relax IEEE, 2017, pp. 3246--3250.

\bibitem{el2011survey}
M.~El~Ayadi, M.~S. Kamel, and F.~Karray, ``Survey on speech emotion
  recognition: Features, classification schemes, and databases,'' \emph{Pattern
  Recognition}, vol.~44, no.~3, pp. 572--587, 2011.

\bibitem{davis1980comparison}
S.~Davis and P.~Mermelstein, ``Comparison of parametric representations for
  monosyllabic word recognition in continuously spoken sentences,'' \emph{IEEE
  transactions on acoustics, speech, and signal processing}, vol.~28, no.~4,
  pp. 357--366, 1980.

\bibitem{qin2006new}
S.~Qin and Y.~M. Zhong, ``A new envelope algorithm of {Hilbert--Huang}
  transform,'' \emph{Mechanical systems and signal processing}, vol.~20, no.~8,
  pp. 1941--1952, 2006.

\bibitem{hermansky1991rasta}
H.~Hermansky, N.~Morgan, A.~Bayya, and P.~Kohn, ``Rasta-plp speech analysis,''
  in \emph{Proc. IEEE Int’l Conf. Acoustics, speech and signal processing},
  vol.~1.\hskip 1em plus 0.5em minus 0.4em\relax Citeseer, 1991, pp. 121--124.

\bibitem{wang2005ideal}
D.~Wang, ``On ideal binary mask as the computational goal of auditory scene
  analysis,'' in \emph{Speech separation by humans and machines}.\hskip 1em
  plus 0.5em minus 0.4em\relax Springer, 2005, pp. 181--197.

\bibitem{wang2015speech}
K.~Wang, N.~An, B.~N. Li, Y.~Zhang, and L.~Li, ``Speech emotion recognition
  using fourier parameters,'' \emph{IEEE Transactions on Affective Computing},
  vol.~6, no.~1, pp. 69--75, 2015.

\bibitem{qassim2018compressed}
H.~Qassim, A.~Verma, and D.~Feinzimer, ``Compressed residual-{VGG16} {CNN}
  model for big data places image recognition,'' in \emph{2018 IEEE 8th Annual
  Computing and Communication Workshop and Conference (CCWC)}.\hskip 1em plus
  0.5em minus 0.4em\relax IEEE, 2018, pp. 169--175.

\bibitem{beckmann2019speech}
P.~Beckmann, M.~Kegler, H.~Saltini, and M.~Cernak, ``Speech-vgg: A deep feature
  extractor for speech processing,'' \emph{arXiv preprint arXiv:1910.09909},
  2019.

\bibitem{hansen1997getting}
J.~H. Hansen and S.~E. Bou-Ghazale, ``Getting started with {SUSAS}: A speech
  under simulated and actual stress database,'' in \emph{Fifth European
  Conference on Speech Communication and Technology}, 1997.

\bibitem{livingstone2018ryerson}
S.~R. Livingstone and F.~A. Russo, ``The ryerson audio-visual database of
  emotional speech and song {(RAVDESS)}: A dynamic, multimodal set of facial
  and vocal expressions in north american english,'' \emph{PloS one}, vol.~13,
  no.~5, p. e0196391, 2018.

\bibitem{zhang2018improved}
Z.~Zhang, ``Improved adam optimizer for deep neural networks,'' in \emph{2018
  IEEE/ACM 26th International Symposium on Quality of Service (IWQoS)}.\hskip
  1em plus 0.5em minus 0.4em\relax IEEE, 2018, pp. 1--2.

\bibitem{sokolova2006beyond}
M.~Sokolova, N.~Japkowicz, and S.~Szpakowicz, ``Beyond accuracy, {F1}-score and
  {ROC}: a family of discriminant measures for performance evaluation,'' in
  \emph{Australasian joint conference on artificial intelligence}.\hskip 1em
  plus 0.5em minus 0.4em\relax Springer, 2006, pp. 1015--1021.

\end{thebibliography}

\end{document}